\documentclass{article}
\usepackage{amssymb}
\usepackage{amsmath}
\usepackage[dvips]{graphicx}

\title{Reflection of nanoparticles}

\date{}

\author{$^{a}$M.A.Ratner, $^{b}$A. V.Tur, $^{a, c}$V. V. Yanovsky}

\begin{document}

 \maketitle

$^{a}$\textit{Institute  for Single Crystals NASU, 60 Lenin Ave, 61001 Kharkiv, Ukraine}

$^{b}$\textit{Institut de Recherche en Astrophysique et Plan\'{e}tologie C.N.R.S.-U.P.S., 9, avenue Colonel-Roche 31028 Toulouse Cedex 4, France.}

$^c$\textit{Kharkiv National University, pl. Svobody 4, Kharkiv, 61022 Ukraine}

\abstract{This work is devoted to molecular dynamics modeling of collision of nanoparticle having a small number of degrees of freedom with a structureless plain. The new regularities are established that determine properties of such particles. Generalized collision law is obtained where particle properties are determined by two coefficient, on of which corresponds to restitution  coefficient. The discovered regularity predicts the existence of anomalous mode of particle reflection from a massive plain. In this mode, velocity of nanoparticle after reflection from a plain  can exceed the initial one. The criterion of  realization of such  mode is obtained. Anomalous collision mode was observed during numerical modeling. Physical mechanism are discussed of phenomena that are observed during numerical experiments. }


\section{Introduction}

Interest in the properties of particles with a small number of degrees of freedom has been growing recently due to intensive researches of nanoobjects (see e.g. \cite{ar1}). It is clear, that, with a diminishing particle size, the number of degrees of freedom decreases. One of the general properties of physical bodies concerns regularities of their collision with a structureless plain or with each other.  Thus, it is important to investigate the influence of small number of internal degrees of freedom on the pattern of such collisions. A simplest model for the investigation of collision with a plane of a nanoparticle with a small number of degrees of freedom was proposed in the work \cite{ar2}. Such composite particle consists of a shell with several particles inside it \cite{ar2}. The shell and the internal particles can move along a chosen direction only, colliding with each other absolutely elastically.  In other words, internal degrees of freedom interact with each other and with a shell absolutely elastically and energy dissipation is not taken into account.  Despite of the simplicity of the proposed model, the obtained results are quite general and, in a certain sense, relate also to scattering of more complex particles with a small number of degrees of freedom.  It is worth to note, that such composite particles have been realized also experimentally.  The example of such structurally complex particles are molecules of rotaxanes \cite{ar3} and nanotube peapods \cite{ar4}, \cite{ar5}.

For macroscopic bodies with a very large number of internal degrees of freedom, the collision with a structureless plane is described by the relation $v_2=-e v_1$. Here $v_1$ is particle velocity prior to its collision with a plane, $v_2$ is its velocity after the collision, $e$ is the restitution coefficient, that is determined by the properties of particle material and does not depend on particle velocity. For actual macroscopic particles, the relation $0 < e < 1$ is always valid. In other words, particle velocity after the collision with an immobile plane can only diminish.The existence of another possibility was discovered in the work \cite{ar2} due to using a simple model. It turned out, that for the collision with a plane of a composite particle with a small number of degrees of freedom, the unusual regimes can be observed.

In the given work, the collision with a plane of more realistic particles with a small number of degrees of freedom is investigated. Using the molecular dynamics method, the particle, consisting from a relatively small number of atoms was modeled. Molecular dynamics methods have now found wide range of applications in physics and chemistry of solid matter \cite{ar6} and for modeling clusters consisting of several to several thousands atoms (see e.g. \cite{ar7}, \cite{ar8}). In the given work, atoms, constituting the particle, interact via Lennard-Jones potential. The collision of such particle with a structureless plane was modeled. The collision velocities of nanoparticle were restricted to nondestructing ones. As a result, new regularities of collision of such particles with a plane were established. In particular, the generalized collision law was obtained, where particle properties are, by this time, determined by two coefficients, one of which corresponds to restitution coefficient while another can be naturally called transformation coefficient.  The revealed regularity predicts the existence of anomalous mode of nanoparticle scattering from a massive plane. In this mode, the velocity of a reflected particle exceeds  that of an incident one. At the same time, conservation laws are obeyed with a high degree of precision.  The existence of such unusual models was also established as a result of collision modeling.

\textbf{}

\section{The purpose of the work and formulation of the problem}

\begin{figure}
  \centering
  \includegraphics[width=6 cm]{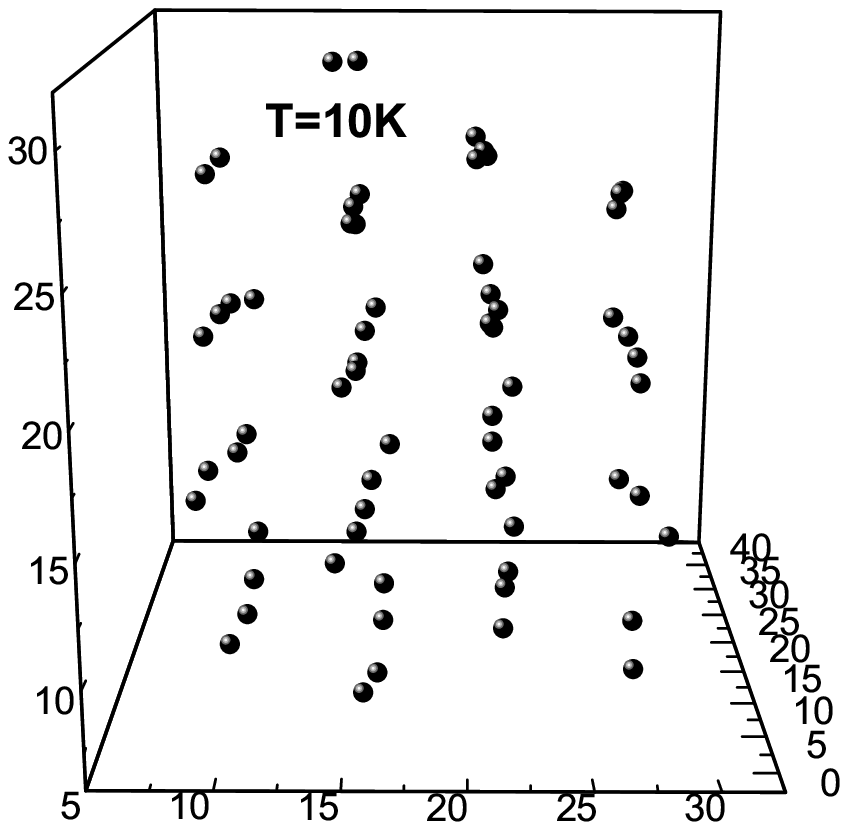}\
  \includegraphics[width=6 cm]{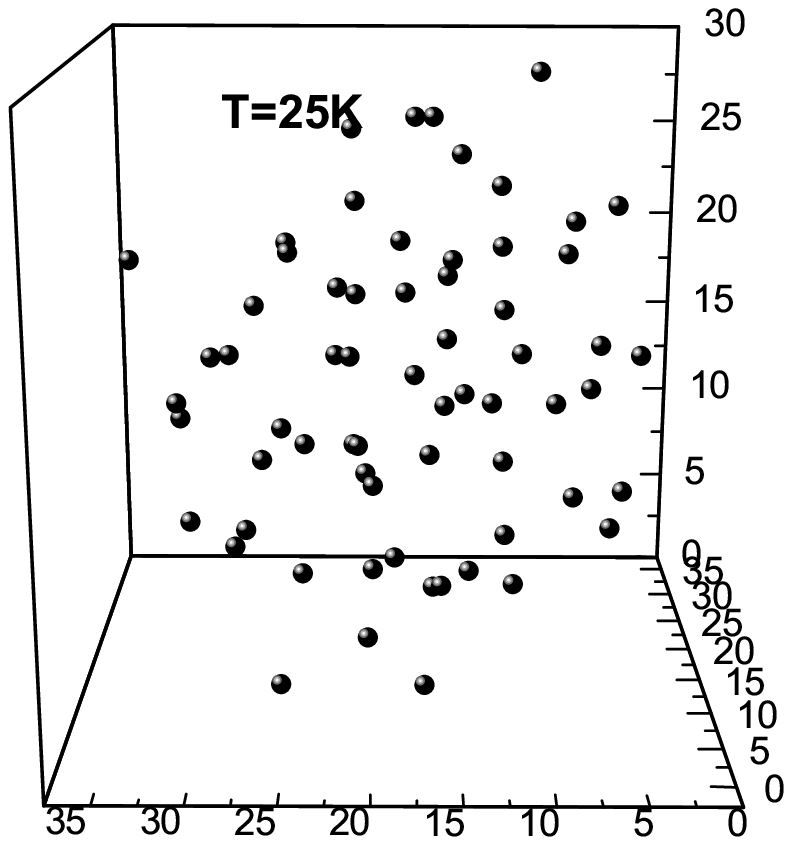}\\
  \caption{he positions of nanoparticle atoms in a chosen moment of time are shown, in the left figure (a) temperature is $T=10 K$, in the right one (b) $T=25 K$}\label{fg1}
\end{figure}

In the given work, the main regularities are considered of the collision of structurally complex nanoparticle with a structureless plane. The collision velocities of nanoparticles are restricted to nondestructing ones. Initially, nanoparticle atoms were placed in the cubic lattice nodes with interatomic distances $a = 4,816$ {\AA}. In the model, the number $N$ of atoms with the mass $m=39,9$ $a.u.$ $=3.5•10^{-30}$ kg, that formed the nanoparticles, was chosen equal to $N=64$ and $N=512$.  Atoms were interacting according to Lennard-Jones potential with the following parameters, taken for Argon atoms: $\sigma =3,405$ {\AA}, $\varepsilon =0,0104$ $eV$. On the first stage of calculations, the system is relaxing until it reaches $2 K$ temperature during $20$ $ps$. Then, heating of the system up to a certain temperature value is realized. All atoms of the nanoparticle are given random displacements, then, system is relaxing again during 20 $ps$. The characteristic collision time between nanoparticle atoms is estimated as $\tau = \sigma (\frac{m}{\varepsilon})^{1/2} \sim 5,16·10^{-14}$ $s$, so that chosen relaxation time is large enough for the system to reach equilibrium. The displacement range was fitted in such a way that system reached a given temperature ($T=10K$ or $25K$) after the relaxation. The temperature value of the relaxed system was averaged over the time during 5 $ps$. It was checked that average temperature value does not change neither with increase of averaging time interval nor with increase of relaxation time. Naturally, the initial crystal structure is changing depending on the temperature value. Thus, at lower temperatures  ($T=10K$) it is preserved, while at higher ones ($T=25K$) it becomes hardly noticeable (see Fig.\ref{fg1}). The equations of motion are solved numerically via velocity Verlet algorithm \cite{ar10} with velocities \cite{ar11} with a time step $dt = 10^{-6} \div 10^{-5} \tau $. The temperatures, chosen for modeling, are lower than melting temperature of Argon  $T=83.4 K$, that  provides for nanoparticle stability.

For a given nanoparticle, consisting of $N$ atoms, the number $Q$ of random realizations with the given temperature $T$ is created.  For particles, containing $N=64$ atoms, $Q=50$, while for those, containing $N=512$ atoms, $Q=25$. For each random realization, the collision with a structureless plain is modeled. To impart the directed motion to the particle, all its atoms are given at the same moment of time the addition to the velocity $v_1$ in the normal to the plane direction. (It is checked beforehand, that the initial center of mass velocity vector components equal to zero.) After that, the numerical integration of motion equations is conducted by the described above method. As soon as any of nanoparticle atoms reaches the plain, its velocity components are instantly changed to their mirror images, then motion equation integrating is conducted in the previous mode. After 20 $ps$ since the collision, the system temperature is measured via the described above procedure. Nanoparticle initial velocities vary in the range from $25$ to $220$ $m/s$. Such velocities do not cause destruction of the nanoparticle.

\section{The numerical results}

Let us begin with investigating the dependence of the reflection velocity of nanoparticle, $v_2$, on its initial velocity \textit{v}1. In the case of macroscopic bodies, such dependence is simple enough: $v_2=-e v_1$ and is reduced to the linear law. Here $e$ is restitution coefficient, that is determined by the material properties and does not depend on the initial velocity of the body, colliding with the plane. This low is known from the Newton time. Defining the effective restitution coefficient as $e_{ef} =|v_2 / v_1 |$, let us consider its dependence on the dimensionless velocity of nanoparticle direct motion in the normal to the plane direction.
\begin{equation}\label{eq1}
    q=\frac{v_{1} }{v_{T} }
\end{equation}
Here $v_T$ is characteristic velocity of thermal motion of nanoparticle atoms, and is defined as $v_T =(3k_b T/ m )^{1/2}$,  $k_b=1.3806488(13) \times 10^{-23}$ $J/K$ is Boltzmann constant. All the dependencies, presented below, are averaged over $Q$ random realizations and temperatures $T=10$ $K$ and 25 $K$.
\begin{figure}
  centering
  \includegraphics[width=7 cm]{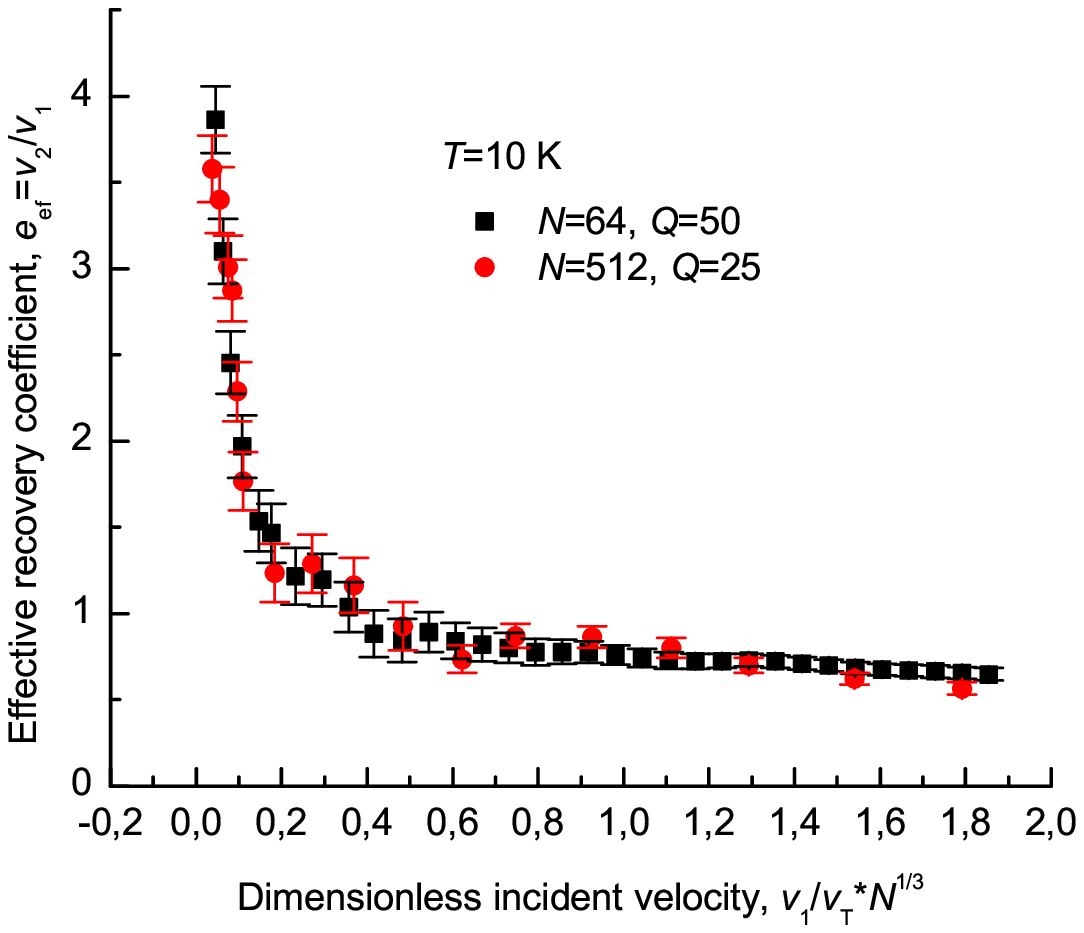}\
  \includegraphics[width=7 cm]{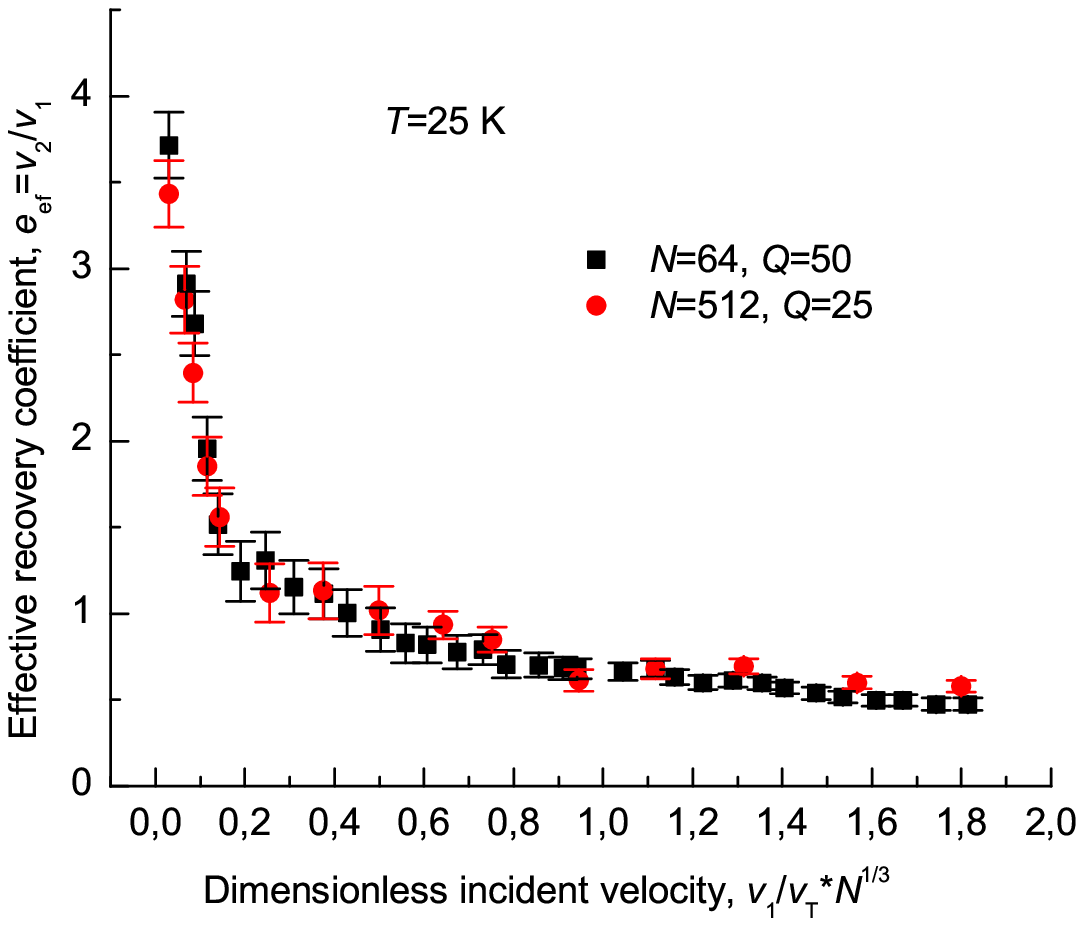}\\
  \caption{The effective restitution coefficient $e_{ef}=|v_2 / v_1|$ as a function of ratio of incident velocity to the thermal one, $q=v_1 / v_T$ and of number of atoms in nanoparticle, $N$. In the left figure (a), temperature is $T=10 K$, data, corresponding to nanoparticle, containing $N=64$ atoms are indicated by squares; data, relating to $N=512$, are indicated by circles. The vertical segments corresponds to calculating error. In the right figure (b), similar dependencies are shown at temperature $T=25 K$.}\label{fg2}
\end{figure}

Dependence of effective restitution coefficient $e_{ef}$ on $q N^{1/3}$ is presented in Fig.\ref{fg2}. Multiplier $N^{1/3}$ is introduced in order to demonstrate the universal dependence of the effective restitution coefficient on the number of atoms, constituting nanoparticle. First of all, the obtained dependence $e_{ef}$ on $q N^{1/3}$ is close to the inversely proportional one.
\begin{equation}\label{eq2}
    \frac{v_2}{v_1}=e + const \frac{v_T}{N^{1/3}v_1}
\end{equation}
where $e$ and $const$ are some constants. This implies, that introduced in such a way restitution coefficient depends not only on the material properties, but also on the velocity of incident particle. This is the principal difference from the independence of \textit{e} from the velocity of an incident structureless macroscopic body. In other words, such coefficient does not fit as  material characteristic that determines the collision law of nanoparticle with a structureless plane.  In principle, this leads to the necessity of introducing another collision law for nanoparticles or particles with a small number of degrees of freedom.

Besides this, the obtained dependence of effective restitution coefficient demonstrates one more unusual property, that cannot be realized for the collision with a plain of macroscopic structureless bodies. Namely, for macroscopic bodies, the reflected velocity can be only less than the incident one. For nanoparticles, as it follows from the modeling results, at $ q N^{1/3} \ll 1$, velocity of a particle after its collision  with the plane is larger by an absolute value than that before collision ($e_{ef} > 1$). Of course, conservation laws are not violated in this case. Such reflection regime is accompanied by cooling of internal degrees of freedom of the nanoparticle. With an increase of \textit{q}, the reflected velocity becomes smaller by an absolute value than the initial one and decreases slowly ($e_{ef} < 1$) which is accompanied by heating of the nanoparticle. At the same time, the total energy of the particle is preserved during modeling within the accuracy of $10^{-6} \%$). Thus, unlike macroscopic particles, for a model nanoparticle, the reflected velocity can exceed the incident one. The performed modeling gives the critical value $q_{cr}$, at which  transition occurs from $e_{ef} > 1 $ to  $e_{ef} < 1$
\begin{figure}
  \centering
  \includegraphics[width=6 cm]{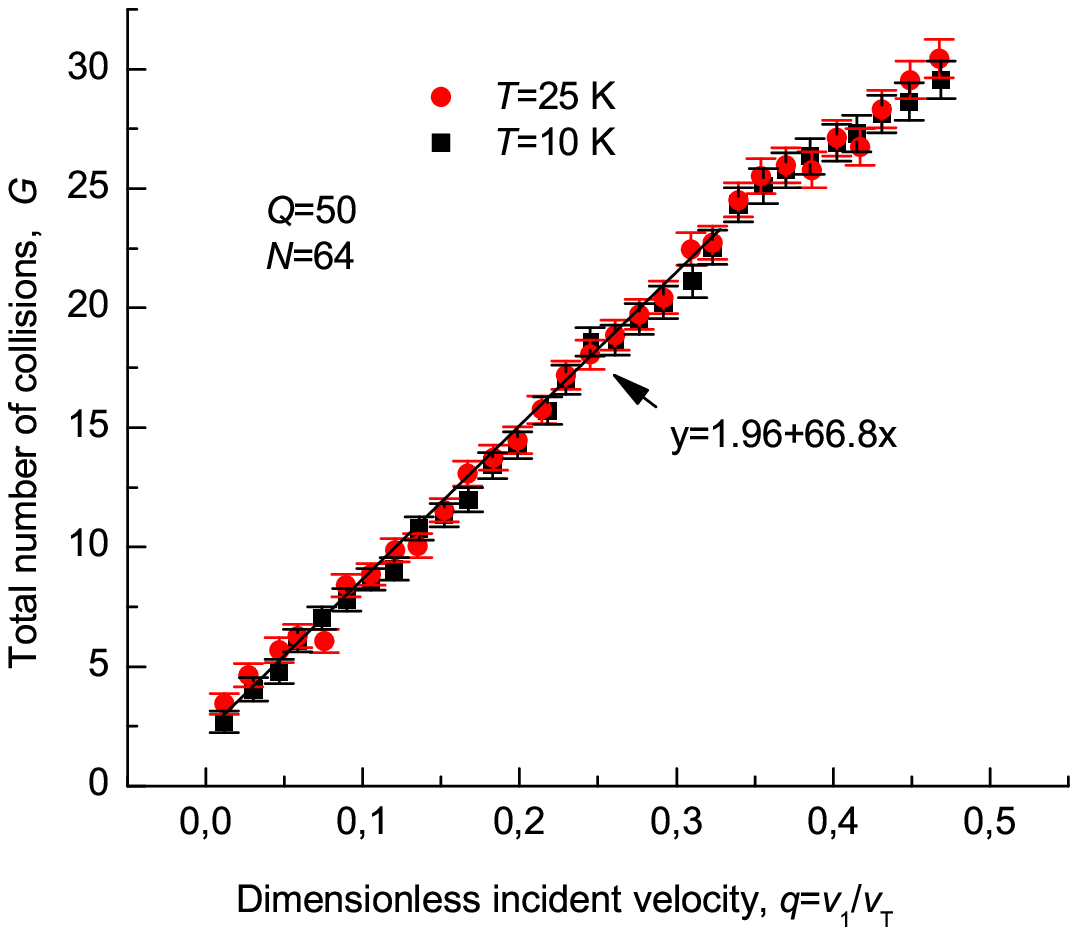}\
  \includegraphics[width=6 cm]{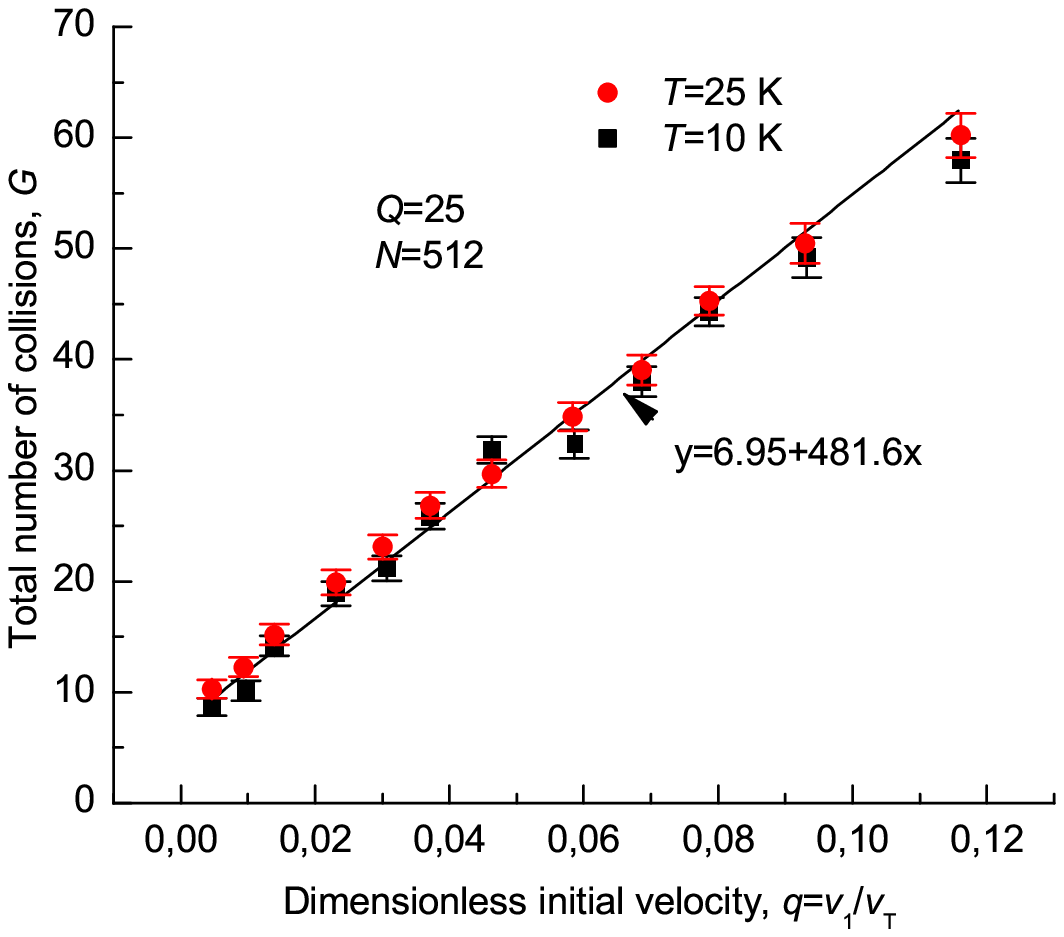}\\
  \caption{Number of collisions, $G$, of nanoparticle atoms with the plain dependent on ratio of incident velocity to the thermal one, $q=v_1 / v_T$. Data in the left figure (a) correspond to the nanoparticle, containing \textit{N}=64 atoms. Data, relating to the temperature $T=10 K$ are indicated by squares while those relating to $T=25 K$ are indicated by circles. In the right figure (b), similar dependencies are shown for $N=512$.  Linear dependence is observed in a wide velocity range $0 < v_1 < v_{T}/2$.}\label{fg3}
\end{figure}

\begin{equation}\label{eq3}
    q_{cr} \approx \frac{1}{2 N^{1/3}}
\end{equation}

It is clear, that for macroscopic bodies with a gigantic number of internal degrees of freedom, $N \to \infty$ and $q_{cr}>0$, so that it is impossible to observe the described phenomena. It is worth to note, that, as it follows from Fig.\ref{fg2}, the form of dependence of $e_{ef}$ on $q$, including the critical value $q_{cr}$, is virtually non dependence on size and temperature of nanoparticles in the investigated range of $N$ and $T$.

One more interesting characteristics of the process is $G$, the full number of collisions of nanoparticle atoms with the plane during their interaction. Numerical data on the dependence of $G$ on $q$ are presented in Fig.\ref{fg3} for the temperatures $T=10 K$ and $25 K$. This dependence is linear with a proportionality coefficient $k$ close to $N$. Numerical modeling gives $k=66.8$ for $N=64$ and $k=481.6$ for $N=512$.

\begin{figure}
  \centering
  \includegraphics[width=6 cm]{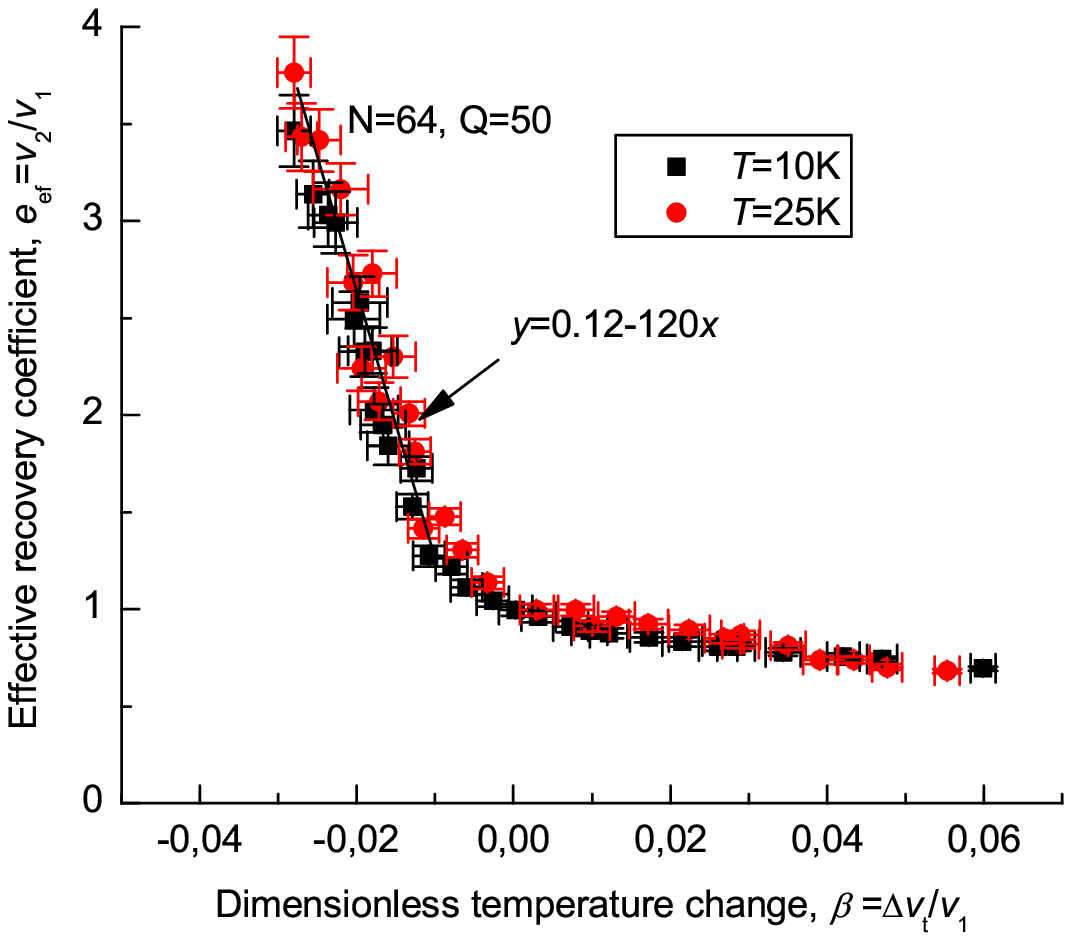}\
  \includegraphics[width=6 cm]{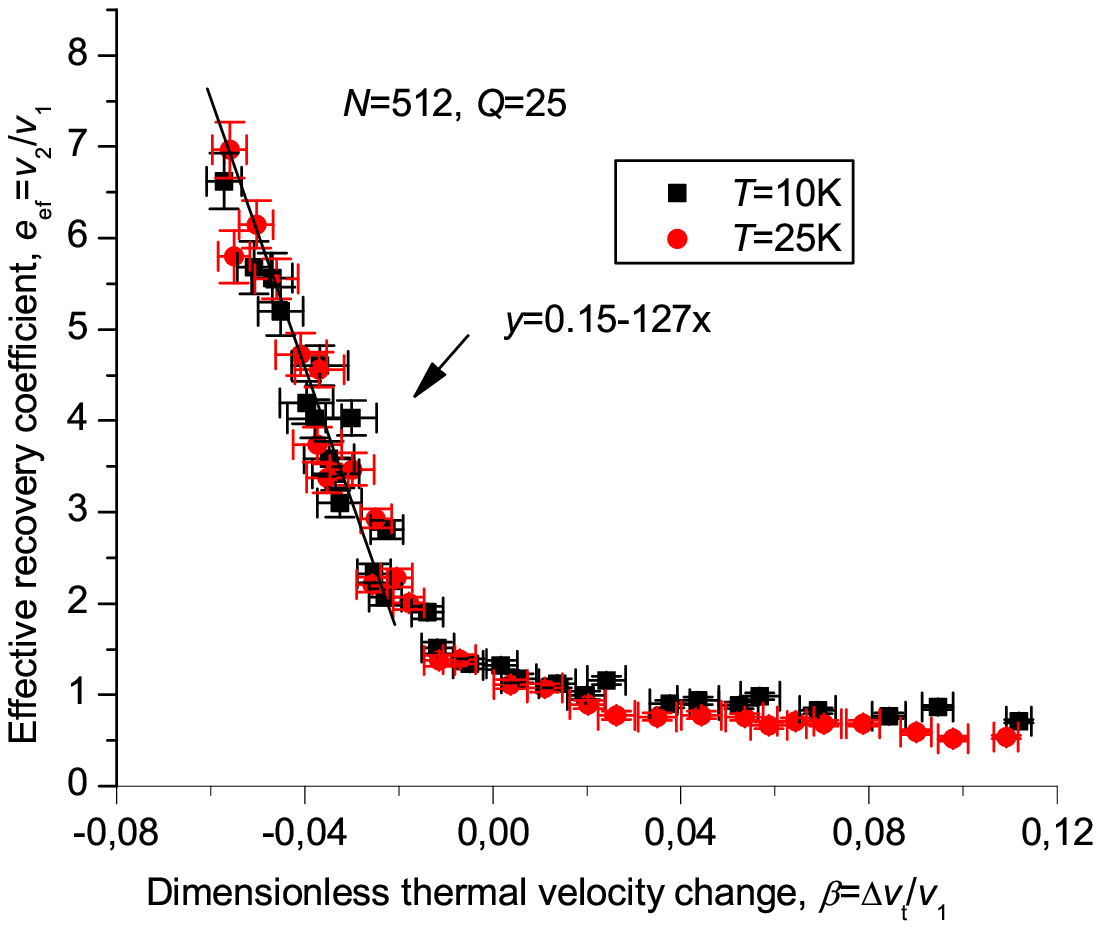}\\
  \caption{Dependence of the effective restitution coefficient $e_{ef}= v_2 / v_1$ on the value of $\beta = v_T / v_1$. Data in the left figure (a) correspond to the nanoparticle, containing $N=64$ atoms. Data, relating to the temperature $T=10 K$ are indicated by squares while those relating to $T=25 K$ are indicated by circles. In the right figure (b), similar dependencies are shown for $N=512$.}\label{fg4}
\end{figure}

Reverting to establishing the reflection law for a particle with a small number of internal degrees of freedom, let us analyze the dependence of $e_{ef} =  v_2 / v_1$ on the ratio of mean thermal velocity change of nanoparticle atoms, $\Delta v_T$, to the initial incident velocity,$ v_1$ ($v_T$ is taken with an account of sign). Let us introduce the notation $\beta =\Delta v_T / v_1$. The results of numerical calculations are presented in Fig.\ref{fg4}. It is easy to notice from the figure that, in a wide range of  $\beta $ values, the linear dependence is observed
\[\frac{v_{2} }{v_{1} } =e+k \beta \]
with proportionality coefficients $k \approx 120$ for $N=64$ and $k \approx 127$  for $N=512$. It is interesting to note, that these values are practically independent on $N$. In principle, for $\beta > 0$, the dependence is also close to linear one, but with different values of proportionality coefficient. It is important to emphasize that coefficient $k$ value does not depend on the particle temperature. Thus, its value is preserved at temperatures $T=10 K$ and $T=25 K$ within calculating error.

Thus, the generalization of the collision law for the bodies with a small number of degrees of freedom takes on a simple form
\begin{equation}\label{eq4}
    v_{2} =e v_{1} +k \Delta v_{T}
\end{equation}
Now, two constants, $e$ and $k$, that do not depend on the incident particle velocity, are present in this law. In a certain sense, these constants depend only on the material properties of the nanoparticle and the wall. It is naturally to preserve the notation of restitution coefficient for $e$. Coefficient $k$, that characterizes nanoparticle properties, can be termed as transformation coefficient. It is easy to note, that from this collision law follow the dependencies, presented in Fig.\ref{fg2},  as well as the exact reversal proportionality between $e_{ef}= e +k \Delta {v}_T / v_1$ and $q$. It is worth to note that coefficient $k$ value depends on the sign of $\beta $. For positive $\beta $ this coefficient takes on smaller values and depends in a differen way on the number of atoms. Preliminary data indicate the inversely proportional dependence on $N^{1/3} $. Thus, value of this coefficient depends on the collision mode.

\section{Discussion}

Let us now discuss regularities obtained as a result of numerical calculations. We will use simple enough consideration in order to explain them. At first, we will concentrate on the anomalous reflection region where $v_T \gg v_1$. At such ratio of direct and chaotic (thermal) components, one can expect that, during collision of a particle with a wall, the momentum transfer $P_1 = m N v_1$ will be realized by collisions of the atoms with thermal velocities directed towards the wall. In a certain sense, this remains some peculiar Fermi acceleration. Lets us suppose that, after collisions of such "hot" atoms with a wall, particle momentum changed to the opposite one, and the particle reflected from the obstacle. Then, the reflected particle momentum becomes equal
\[P_2 = -mG (v_1 +v_T)\]
It is natural to suppose  that the rest of the kinetic energy of direct motion transforms to the energy of chaotic (thermal) motion and does not contribute to the total momentum of the particle. Thus, this is the sum of averaged momentums of atoms that constitutes the momentum of reflected particle that equals to
\[P_2 =-m N v_2\]
Comparing these two expressions, we obtain the expression for the velocity $v_2$ of reflected particle
\[N v_2 = G (v_1 + v_T )\]
or, for the ratio of velocities, we obtain
\begin{equation}\label{eq5}
    \frac{v_2}{v_1} = \frac{G}{N}+\frac{G v_T}{N v_1}
\end{equation}
Thus, in the region $v_T \gg v_1$ one can expect hyperbolic dependence of $v_2 / v_1$ on $q$, like that obtained by numerical modeling. Certainly, these rough considerations can be made more detailed via tedious averaging over velocity distribution function. However, it is important to note, that, even from such simple considerations, follows the mechanism of particle reflection with a velocity exceeding the incident one.  This mechanism is realized in the velocity region $v_T \gg v_1$. Just this type of anomalous reflection modes was observed during numerical modeling.

Thus, it is left for us to discuss the number $G$ of collisions of atoms with a massive wall during the total time of particle reflection. Of course, for the estimation of $G$, we, as before, consider region $v_T \gg v_1$. In order to estimate the collision number, we use the fact, that the atoms that lie on one side of a particle, collide with a wall even at negligibly small velocity $v_1$. The number of atoms on one side can be estimated as $N^{2/3}$. Moreover, it is clear that the greater is the incident velocity, the larger is the collision number required to change momentum $m N v_1$ to the opposite. Taking into consideration that each collision of an atom with the wall transfers to the latter the momentum equal approximately to $m v_T$, let us write down the general formulae in the following form:
\begin{equation}\label{eq6}
    G \approx \frac{N^{2/3}}{6}+N \frac{v_1}{v_T}
\end{equation}
The obtained dependence is in a good agreement with numerical modeling data (see Fig.\ref{fg3}). Thus, as can be seen from the left part of the figure, for $N = 64$, linear coefficient at $v_T / v_1$ is close to $N$, as well as for $N = 512$ in the right figure. The additive constant amounts to 1.96  for $N = 64$ and 6.95 for $N = 512$.  The Eq.\eqref{eq6} gives the additive constant equal to 2.67 for $N = 64$ and 10.67 for for $N = 512$, that is quite close to the numerical data. This correspondence becomes even better if one takes into account that dividing by 6 gives overestimated number of atoms on the side and does not allow for the deviation of particle shape from a cube. Division by 8 gives much better coincidence of additive constants with numerical data, namely, for $N = 64$  it equals 2, while for $N = 512$ it equals 8. Besides of this, it is easily to note, that such dependence of $G$ leads to the universal dependence of $v_1 / v_2$ on parameter $N^{1/3} q$ , that is demonstrated in Fig.\ref{fg2}. Let us turn now to discussing the realization criterion of anomalous reflection mode. Returning to the Eq.\eqref{eq5}, such condition can be easily obtained:
\[ \frac{G v_T}{N v_1} \geq 1\]
Here, the condition $v_T / v_1 \gg 1$, at which anomalous regime is observed, is taken into account. Let us recall, that collision number $G \sim N^{2/3}$ is determined by surface atoms.  Thus, the condition of realization anomalous regime takes on the following form
\[\frac{v_1}{v_T}N^{1/3} \leq 1\]
that is in a good agreement with the relation, obtained by numerical modeling \eqref{eq3}.  For estimation by  an order of magnitude, the difference between the right parts of inequalities is nonsignificant. If one defines the critical value of the ratio between incident and thermal velocities, that separates different regimes, as
\begin{equation}\label{eq7}
    q_{cr} \approx \frac{1}{2 N^{1/3}}
\end{equation}
it is possible to compare it with $q_{cr}$, obtained directly from numerical modeling. Namely, for  $N = 64$, the numerical experiment gives $q_{cr} = 0.11$, and  Eq.\eqref{eq7} gives $q_{cr} \approx 0.13$, while for $N = 512$ numerical modeling gives $q_{cr} = 0.06$, while from Eq.\eqref{eq7} follows the value $q_{cr} = 0.07$ (see results, shown in Fig.\ref{fg2}).

Let us discuss now linear dependence of ratio $v_1 / v_2$ from $\beta = \Delta v_T / v_1$ (see Eq.(\ref{eq4})).  It is easy enough to predict such dependence in the region $v_T / v_1 \gg 1$.  Indeed, in this region the momentum transfer is realized by hot atoms with velocities  $v \approx v_T$, therefore, thermal velocity change $\Delta v_T$ is proportional to $v_T$. Then, from Eq.\eqref{eq2}, follows linear dependence from $\beta$ like in Eq.\eqref{eq4}.

The last thing to discuss is nanoparticle reflection in the velocity region $v_1 \geq v_T$. In this region, particle reflection pattern is more traditional. Indeed, in this region the momentum transfer is realized by atoms with velocities of the order of $v_1$. Therefore, momentum after reflection sums up from atom momentums
\[P_2 = -mG v_1\]
and determines momentum of reflected particle
\[P_2 = -m N v_2\]
Equating those two expressions, one obtains
\[v_2 = \frac{G}{N} v_1\]
Now, via using the estimate for collision number \eqref{eq6}, but in another limiting case $v_1 \geq v_T$, it is easy to obtain linear dependence
\[\frac{v_2}{v_1} \sim \frac{v_1}{v_T}\]
Such linear dependence does not contradict to experimental data, shown in Fig.\ref{fg2}.

\section{Conclusions}

Let us summarize briefly modeling results. If a particle is reflecting from a structureless plane, for initial velocities of the particle much smaller than velocity of thermal motion of its atoms, the effective restitution coefficient $e_{ef}$ can exceed unity. The cause of this phenomena lies in the change of relation that determine particle velocity after its collision with the wall. This new relation is obtained in the present work. The collisional properties of the particles are determined now by two material characteristics, namely by restitution coefficient $e$, that is always smaller than unity, and by transformation coefficient. With increasing number of atoms that constitute nanoparticle, the anomalous mode disappears. In other words, anomalous collisional properties are inherent to nanoparticles only, whereas for macroscopic bodies such behavior is impossible. The condition of realization $e_{ef} > 1$  is determined by inequality $q_{cr} < 1/(2 N^{1/3})$.

\end{document}